\pdfoutput=1
\documentclass[sigplan,10pt,nonacm]{acmart}

\setcopyright{none}
\renewcommand\footnotetextcopyrightpermission[1]{}

\acmConference[WoAIS '26]{Workshop on AI and Serverless}{December 14-18, 2026}{Tarragona, Spain}
\acmYear{2026}

\usepackage[capitalise,noabbrev]{cleveref}
\usepackage{tikz}
\usepackage{tcolorbox}
\usepackage{xspace}

\usetikzlibrary{arrows.meta}
\usepackage{fontawesome5}%
\colorlet{figtext}{black}%

\definecolor{findingbg}{HTML}{E9EEF7}
\definecolor{findingframe}{HTML}{A7B7D3}
\newcounter{finding}
\newtcolorbox{findingbox}{colback=findingbg, colframe=findingframe, boxrule=0.4pt, arc=0pt,
  left=3pt, right=3pt, top=2pt, bottom=2pt, before skip=4pt, after skip=4pt, fontupper=\itshape}
\newcommand{\finding}[2]{\refstepcounter{finding}\label{#1}%
  \begin{findingbox}\textup{\textbf{Finding~\thefinding:}} #2\end{findingbox}}
\newcommand{\reffinding}[1]{Finding~\ref{#1}}

\DeclareRobustCommand{\circled}[1]{%
    \tikz[baseline=(char.base)]{\node[shape=circle, fill=black, text=white, inner sep=1pt] (char) {#1};}%
}

\newcommand{\dataset}{\texttt{LLMFineTuningBench}\xspace}

\newif\ifanon \anonfalse
\newcommand{\kavier}{\ifanon \texttt{simulator} (\S\ref{sec:conclusion})\else Kavier\fi}
\newcommand{\coastline}{\ifanon \texttt{recommender}\else Coastline\fi}

\begin{document}

\title{LLM Fine-Tuning Workloads: Characterization, Challenges, and Opportunities for Serverless AI}
\subtitle{Technical Report}
\titlenote{Technical report of the article published at Middleware WoAIS '26~\cite{nicolae26woais}.}

\author{Radu Nicolae}
\orcid{0009-0007-0318-9266}
\affiliation{\institution{Vrije Universiteit Amsterdam}\city{Amsterdam}\country{The Netherlands}}
\email{r.nicolae@vu.nl}

\author{Daniele Lotito}
\orcid{0000-0003-0320-0043}
\affiliation{\institution{IBM Research Europe}\city{Dublin}\country{Ireland}}
\email{daniele.lotito@ibm.com}

\author{Alexandru Iosup}
\orcid{0000-0001-8030-9398}
\affiliation{\institution{Vrije Universiteit Amsterdam}\city{Amsterdam}\country{The Netherlands}}
\email{a.iosup@vu.nl}

\begin{abstract}
Serverless execution of Large Language Models~(LLMs) is becoming important, reflecting their wide societal use. 
An increasingly present stage in LLM pipelines is fine-tuning, which specializes trained models and improves inference performance. 
Serverless operation of LLM fine-tuning workloads shifts operational concerns from the user to the provider. 
For these new workloads, providers %
face %
important
operational complexities, such as performance issues, resource waste, and even service failure due to misconfiguration. %
However, optimizing such workloads is still at an early stage, and, relevant to this work, relatively little is currently known about how LLM fine-tuning workloads run in practice. %
In this work, we address the knowledge gap by characterizing the execution of LLM fine-tuning workloads on clusters equipped with different GPUs, running LLM models at various scales. 
We analyze IBM's \dataset, a large-scale, diverse profiling dataset of 30,920 fine-tuning experiments %
that records both successful and failing configurations.
We provide quantitative insight into throughput (expressed in tokens/s), scaling, failure, and potential for performance optimization for four commonly used GPU types, in setups of 1 to 32 GPUs. 
We find that (1) job feasibility~(run or fail) depends on workload properties, %
with failure ratios rising from 52\% for small batch sizes (16-64) up to 95\% for the largest (1,024); 
(2) throughput grows logarithmically with batch size; 
(3) scaling is near-linear when the workload is deployed intra-node, and optimizations such as RoCE substantially improves efficiency for inter-node deployment (e.g., 0.72 vs. 0.26 efficiency without RoCE); %
and (4) system-level optimizations can outweigh application-level ones (RoCE, up to 2.9$\times$, vs. Fast MoE, 1.7$\times$).

\end{abstract}



\maketitle

\section{Introduction}
\label{sec:intro}

Our society and economy are increasingly reliant on AI services, especially Large Language Models~(LLMs), which are increasingly adopted and embedded across academia~\cite{meyer2023chatgpt, alzaabi2023chatgpt, liang2024mapping}, government~\cite{dai2024applications, safaei2024end, qin2024study}, and industry~\cite{DBLP:journals/corr/abs-2408-07326, DBLP:journals/corr/abs-2501-11006}.
Subject to cost and scalability pressures, the typical training-inference two-step pipeline has evolved: 
cheaper and smaller-scale than training, 
and placed in the pipeline in-between training and inference, \textit{LLM fine-tuning} is becoming a common approach to specializing LLMs and improving their inference performance. %
As LLM fine-tuning adoption increases, %
serverless execution could bring important performance improvements, efficiency gains, and operational transparency~\cite{DBLP:journals/cacm/KounevHAIFSRC23}.

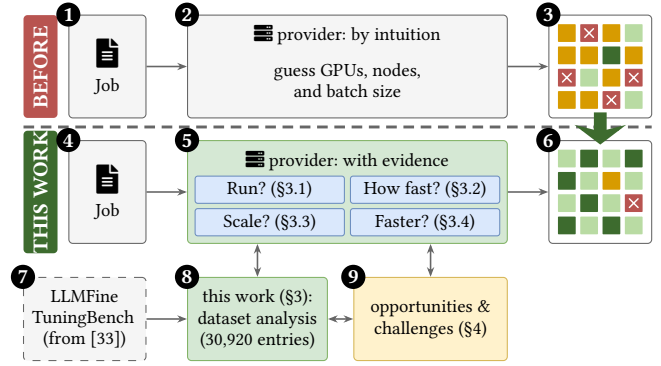
\begin{figure}[t]
  \centering
  \definecolor{dioBlueF}{HTML}{DAE8FC}\definecolor{dioBlueS}{HTML}{6C8EBF}%
\definecolor{dioGreenF}{HTML}{D5E8D4}\definecolor{dioGreenS}{HTML}{82B366}%
\definecolor{dioRedF}{HTML}{F8CECC}\definecolor{dioRedS}{HTML}{B85450}%
\definecolor{dioOrangeF}{HTML}{FFE6CC}\definecolor{dioOrangeS}{HTML}{D79B00}%
\definecolor{dioYellowF}{HTML}{FFF2CC}\definecolor{dioYellowS}{HTML}{D6B656}%
\definecolor{dioGreyF}{HTML}{F5F5F5}\definecolor{dioGreyS}{HTML}{666666}%
\definecolor{dioPurpleF}{HTML}{E1D5E7}\definecolor{dioPurpleS}{HTML}{9673A6}%
\definecolor{dioGreenD}{HTML}{3D6B2E}%
\colorlet{sqD}{dioGreenD}\definecolor{sqG}{HTML}{B9DBA3}\colorlet{sqO}{dioOrangeS}\colorlet{sqR}{dioRedS}%
\begin{tikzpicture}[
  x=1cm, y=1cm, font=\scriptsize, text=figtext,
  box/.style={draw=dioGreyS, rounded corners=1.5pt, align=center, inner sep=0pt,
              minimum height=13.6mm, fill=white},
  ops/.style={box, fill=dioGreyF},
  ev/.style={box, minimum height=7.4mm},
  ibm/.style={ev, fill=dioGreyF, dashed},
  our/.style={ev, draw=dioGreenS, fill=dioGreenF},
  ttl/.style={anchor=base, inner sep=0pt},
  chip/.style={draw=dioBlueS, fill=dioBlueF, rounded corners=1pt, inner sep=0pt,
               minimum width=1.975cm, minimum height=3.6mm},
  badge/.style={rotate=90, text=white, rounded corners=1.5pt, inner sep=0pt,
                minimum width=13.6mm, minimum height=4.5mm, font=\footnotesize\bfseries},%
  >={Stealth[length=1.3mm]},%
  arr/.style={->, semithick, dioGreyS},
  both/.style={<->, semithick, dioGreyS},
  rel/.style={<->, semithick, dioGreyS},
  future/.style={rel},%
  lbl/.style={circle, fill=black, text=white, inner sep=0.9pt, font=\footnotesize\bfseries},%
]
  \def\grid#1#2#3{%
    \foreach \c [count=\i from 0] in {#3} {
      \pgfmathsetmacro{\sx}{#1-0.56+0.295*mod(\i,4)}
      \pgfmathsetmacro{\sy}{#2+0.56-0.295*floor(\i/4)}
      \fill[sq\c, rounded corners=0.3pt] (\sx, \sy) rectangle +(0.235, -0.235);
      \if R\c \draw[white, line width=0.55pt, line cap=round]
        (\sx+0.06, \sy-0.06) -- +(0.115, -0.115) (\sx+0.175, \sy-0.06) -- +(-0.115, -0.115); \fi
    }}

  \node[badge, fill=dioRedS] at (0.225, 0) {BEFORE};
  \node[ops, minimum width=1.02cm] (bj) at (1.11, 0) {{\large\faFile*}\\[2pt]Job};
  \node[ops, minimum width=4.23cm] (bp) at (4.285, 0) {};%
  \node[ttl] at (4.285, 0.35) {\faServer\ provider: by intuition};
  \node[align=center] at (4.285, -0.185) {guess GPUs, nodes,\\and batch size};
  \node[box, minimum width=1.36cm] (bo) at (7.63, 0) {};
  \grid{7.63}{0}{O,R,O,G, O,O,D,O, R,G,O,R, O,O,R,G}

  \draw[dioGreyS, line width=1.1pt, dash pattern=on 4pt off 2.5pt] (0, -0.80) -- (8.31, -0.80);

  \node[badge, fill=dioGreenD] at (0.225, -1.66) {THIS WORK};
  \node[ops, minimum width=1.02cm] (aj) at (1.11, -1.66) {{\large\faFile*}\\[2pt]Job};
  \node[box, draw=dioGreenS, fill=dioGreenF, minimum width=4.23cm] (ap) at (4.285, -1.66) {};
  \node[ttl] at (4.285, -1.31) {\faServer\ provider: with evidence};
  \node[chip] at (3.2575, -1.63) {Run? (\S\ref{sec:analysis:run})};
  \node[chip] at (5.3125, -1.63) {How fast? (\S\ref{sec:analysis:fast})};
  \node[chip] at (3.2575, -2.06) {Scale? (\S\ref{sec:analysis:gpus})};
  \node[chip] at (5.3125, -2.06) {Faster? (\S\ref{sec:analysis:faster})};
  \node[box, minimum width=1.36cm] (ag) at (7.63, -1.66) {};
  \grid{7.63}{-1.66}{G,D,G,D, D,G,O,G, G,D,G,R, D,G,D,G}

  \node[ibm, minimum width=1.62cm, minimum height=11mm] (ds) at (0.81, -3.34) {LLMFine\\TuningBench\, \\(from \cite{lotito26finetuning})};
  \node[our, minimum width=1.85cm, minimum height=11mm] (an) at (3.095, -3.34) {this work (\S\ref{sec:analysis}):\\dataset analysis\\(30,920 entries)};
  \node[ev, draw=dioYellowS, fill=dioYellowF, minimum width=2.03cm, minimum height=11mm] (og) at (5.385, -3.34)
    {opportunities \& \\challenges (\S\ref{sec:insights})};

  \foreach \a/\b in {bj/bp, bp/bo, aj/ap, ap/ag} \draw[arr] (\a) -- (\b);
  \draw[arr] (ds.east |- an.west) -- (an.west);
  \draw[both] (an.north) -- (an.north |- ap.south);
  \draw[future] (og.north) -- (og.north |- ap.south);
  \draw[rel] (an.east) -- (og.west);
  \draw[-{Stealth[length=2.6mm, width=6.8mm]}, line width=4.8pt, dioGreenD]
    ([yshift=0.8mm]bo.south) -- ([yshift=-0.6mm]ag.north);
  \foreach \b/\n in {bj/1, bp/2, bo/3, aj/4, ap/5, ag/6, ds/7, an/8, og/9}
    \node[lbl] at (\b.north west) {\n};
\end{tikzpicture}
  \vspace*{-0.5cm}
  \caption{Fine-tuning as a service configured blindly, based on intuition (\circled{1}-\circled{3}), compared to evidence-based (\circled{4}-\circled{9}). We characterize \dataset~\cite{lotito26finetuning} in \S\ref{sec:analysis} and offer actionable insights in \S\ref{sec:insights}.}
  \Description{Fine-tuning as a service configured blindly, based on intuition, compared to evidence-based. Before: provider by instinct, guess GPUs, nodes, and batch size. This work: provider with evidence, Run? How fast? Scale? Faster? LLMFineTuningBench, this work: dataset analysis (30,920 entries), ongoing work: side-advisor, digital twin, related work.}
  \label{fig:overview}
  \vspace*{-0.3cm}
\end{figure}

Serverless LLM fine-tuning raises known operational concerns, including throughput, scalability, resource use, related efficiency trade-offs, and whether an automatically configured job is feasible (completes or fails). 
This raises the challenge that, as a new workload, LLM fine-tuning has largely unknown execution characteristics. 
Although current characterizations of GPU clusters analyze production traces of deep-learning jobs~\cite{DBLP:conf/usenix/JeonVPQXY19, DBLP:conf/sc/Hu0Y0021, DBLP:conf/nsdi/WengXYWWHLZLD22} and of LLM pretraining and evaluation~\cite{DBLP:conf/nsdi/Hu0WWZC0L0L0024}, only anecdotal evidence exists on LLM \textit{fine-tuning} performance and feasibility~\cite{DBLP:conf/iiswc/XiaKCY0HT24, DBLP:journals/corr/abs-2406-02290}.
Addressing this challenge, in this work we characterize LLM fine-tuning workloads in GPU-based clusters. 

Currently, massive-scale and powerful  LLMs~\cite{DBLP:journals/corr/abs-2108-07258, DBLP:journals/software/LuZXXW24}, such as Llama~\cite{DBLP:journals/corr/abs-2307-09288}, Mistral~\cite{DBLP:journals/corr/abs-2401-04088}, and Granite~\cite{DBLP:journals/corr/abs-2405-04324}, are foundational for LLM operations. However, retraining LLMs from scratch is rapidly pricing out even the largest non-Big Tech enterprises~\cite{ramponi2024llm}, leaving fine-tuning among the few performance- and cost-sustainable techniques for enterprise or smaller-scale specialization~\cite{DBLP:conf/nips/DettmersPHZ23}.
Consequently, platforms such as Amazon Bedrock~\cite{claude-amazon-bedrock}, Databricks Mosaic AI~\cite{databricks-mosaicai}, and IBM's watsonx.ai~\cite{ibm_watsonx_ai} increasingly offer fine-tuning as a service.

The community is working towards
serverless operations for LLM, based on quantitative evidence about how LLM workloads run in practice. 
As illustrated in \Cref{fig:overview} for fine-tuning, 
the
vision is to \textit{replace the current approach to configuring LLM fine-tuning jobs}, which is based on anecdotal evidence and intuition, \textit{with a data-driven approach based on shared profiling information and (as in this work) characterization}. 
Anecdotes and intuition are unlikely to overcome the rapid increase in complexity and unpredictability of fine-tuning workloads~\cite{DBLP:conf/nips/DettmersPHZ23, DBLP:conf/iclr/HuSWALWWC22}.  
As a lesson learned by operators across decades, relying on users to provide accurate estimates of resource use has proven infeasible~\cite{DBLP:conf/jsspp/LeeSHS04}.
As supported by the evidence presented in this work~(\Cref{fig:overview}, component~\circled{8}, detailed in~\Cref{sec:analysis}) jobs configured using these approaches have a high likelihood of failure (boxes marked with $\times$ in~\circled{3}) or high performance degradation (orange boxes). Conversely, changing to an evidence-based approach should alleviate performance and failure~(\circled{6}).
\Cref{sec:insights} discusses opportunities and challenges related to this change.

Configuring LLM jobs is not easy. 
Configurations need to account for diverse GPU types for cost and efficiency reasons~\cite{DBLP:journals/corr/abs-2404-14527}. 
Misconfiguration could trigger worse performance or less efficient use of (GPU) resources. The most important effect, %
job failure, occurs for example when settings lead to out-of-memory crashes~\cite{DBLP:conf/sosp/YinMZZBP11, DBLP:conf/icse/ZhangXZLLY20}. Worse, when one job fails it could, in turn, cause further, correlated failures~\cite{DBLP:conf/europar/GalletYJKIE10,DBLP:conf/grid/YigitbasiGKIE10}. 

Operators of LLM infrastructure are already collecting data about fine-tuning in practice.
IBM released in September 2026 an important dataset~\cite{lotito26finetuning}~(\circled{7}, \Cref{fig:overview}), which is the largest-of-its-kind profiling effort
covering 30,920 fine-tuning experiments across 33 LLMs and 4 commonly used GPU types.
The FAIR~\cite{wilkinson2016fair} dataset \dataset  %
covers the configuration space systematically, via step-wise exploration, and includes failed configurations.
In contrast to production traces, \dataset%
facilitates comprehensive characterization, but records no job arrival patterns.

\textbf{Related work:}
Fine-tuning studies measure a few LLMs on a few GPU types~\cite{DBLP:conf/iiswc/XiaKCY0HT24, DBLP:journals/corr/abs-2406-02290}, and fine-tuning schedulers, e.g., Ymir~\cite{DBLP:conf/ics/GaoZLS0024}, use these measurements as input.
Trace archives for workloads~\cite{DBLP:journals/fgcs/IosupLJADWE08}, failures~\cite{DBLP:conf/ccgrid/KondoJIE10}, 
and workflows~\cite{DBLP:journals/tpds/VersluisMTHPDI20} have proven the value of data collection and characterization, but do not cover LLM fine-tuning.
IBM's AutoConf~\cite{ibm_autoconf} uses \dataset to predict configuration feasibility, but its core ML-model does not improve understanding of configuration effects. 
Lotito et al.~\cite{DBLP:journals/access/LotitoVVPPJ26} propose early ML-models for performance. %
However, no prior work on LLM fine-tuning matches this work's focus on characterization, scope, scale, and diversity.
We make a four-fold contribution:

\begin{enumerate}
    \item We propose an analysis method~(\S\ref{sec:method}), structured around questions and statistical approaches to answer them.

    \item We conduct the analysis and provide a detailed characterization of LLM fine-tuning workloads in practice~(\S\ref{sec:analysis}):
    whether jobs run and the key levers to job feasibility, %
    job performance, %
    and job scalability, %
    and how to make them faster. %

    \item We present actionable insights related to the characterization~(\S\ref{sec:insights}).
    We envision an online recommender system that helps auto-configure LLM fine-tuning workloads as a key part of the serverless infrastructure, and discuss its challenges and opportunities.

    \item Open science: We release the reproducibility capsule for this project, which includes the analytical tools and results: \href{https://github.com/atlarge-research/llm-finetuning-workload-characterization}{\nolinkurl{github.com/atlarge-research/llm-finetuning-workload-characterization}}
    
\end{enumerate}

\section{Dataset Analysis Method} \label{sec:dataset} \label{sec:method}
\vspace*{-0.15cm}

\begin{table}[t]
\caption{Summary of \dataset~\cite{lotito26finetuning}, per GPU type. 
\#Exp. = experiment count; 
\faCheck{} = ratio of jobs that ran successfully, 
\faBan{} = rejected before launch, 
\faTimes{} = failed at runtime. \textbf{Dataset} = entire dataset. 
Rows sum to 100\%.}
\label{tab:background:dataset_stats}
\vspace*{-0.3cm}
\footnotesize
\centering
\setlength{\tabcolsep}{2pt}
\setlength{\aboverulesep}{0pt}\setlength{\belowrulesep}{0pt}%
\renewcommand{\arraystretch}{1.12}
\begin{tabular}{|l|r|r|r|r|r|r|r|r|}
\toprule
\textbf{GPU type} & \textbf{Mem.} & \textbf{\#Exp.} & \textbf{\faCheck $[\%]$} & \textbf{\faBan $[\%]$} & \textbf{\faTimes$[\%]$} & \textbf{LLMs$^{b}$} & \textbf{GPUs} & \textbf{Nodes} \\
\midrule
A100-SXM$^{a}$  & 80\,GB & 27{,}837 & 36.6 & 31.5 & 31.9 & 24 & 1-32 & 1-4 \\
A100-PCIe & 80\,GB &  1{,}818 & 42.2 & 11.3 & 46.5 & 13 & 1-4 & 1 \\
L40S            & 48\,GB &  1{,}068 & 30.8 & 15.5 & 53.7 &  9 & 1-4 & 1 \\
H100-PCIe       & 80\,GB &      197 & 51.3 &  1.0 & 47.7 &  3 & 1-4 & 1 \\
\midrule
\textbf{Dataset} & & \textbf{30{,}920} & \textbf{36.8} & \textbf{29.6} & \textbf{33.6} & \textbf{33} & \textbf{1-32} & \textbf{1-4} \\
\bottomrule
\multicolumn{9}{@{}l@{}}{\scriptsize $^{a}$A100-SXM is the only GPU type with multi-node jobs, RoCE, and GPTQ-LoRA.} \\
\multicolumn{9}{@{}l@{}}{\scriptsize $^{b}$Different combinations of the 33 LLMs tested overall are tested for each GPU type.} \\
\end{tabular}
\end{table}

\begin{table}[t]
\vspace*{-0.35cm}
\caption{Four questions under analysis, by subsection (\S).}
\label{tab:questions}
\vspace*{-0.4cm}
\footnotesize
\centering
\setlength{\tabcolsep}{1.6pt}
\setlength{\aboverulesep}{0pt}\setlength{\belowrulesep}{0pt}%
\renewcommand{\arraystretch}{1.12}
\begin{tabular}{|l|l|l|l|}
\toprule
\textbf{ID} & \textbf{Question (\S)} & \textbf{Metric} & \textbf{Analysis target} \\
\midrule
Q1 & Will the job run? (\ref{sec:analysis:run}) & failure ratio [\%] & per factor \\
Q2 & How fast? (\ref{sec:analysis:fast}) & throughput [tokens/s] & median per batch size \\
Q3 & Will it scale? (\ref{sec:analysis:gpus}) & efficiency & intra- vs.\ inter-node \\
Q4 & How to go faster? (\ref{sec:analysis:faster}) & speedup [$\times$] & matched pairs \\
\bottomrule
\end{tabular}
\vspace*{-0.5cm}
\end{table}

\begin{figure*}[t]
  \centering
  \includegraphics[width=\textwidth]{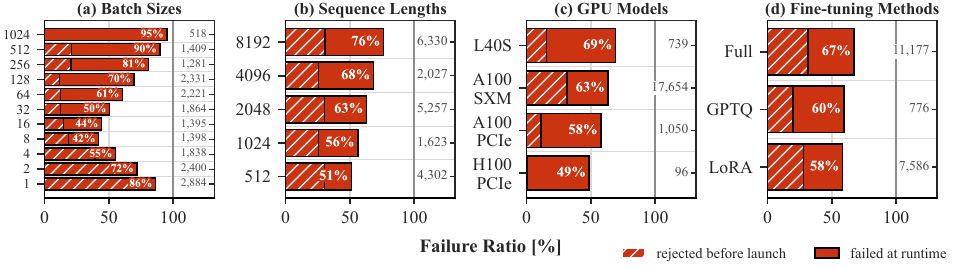}
  \vspace*{-0.7cm}
  \caption{Failure ratios per (a) total batch size, (b) sequence length, (c) GPU model, and (d) fine-tuning method (GPTQ=GPTQ-LoRA). Panels (c) and (d) are sorted by failure ratio. Rows depict the failure ratio (bar and \%) and the number of failing experiments (count), e.g., 69\%~(739) for (c) L40S. Hatched: rejected before launch; solid: failed at runtime.}
  \Description{Sfttrainer: failure ratios per GPU, method, batch size, sequence length.}
  \label{fig:failures}
  \vspace*{-0.3cm}
\end{figure*}

Our method has three stages: 
(\emph{1}) data collection, here, already done by IBM~(\circled{7}, \Cref{fig:overview}),
(\emph{2}) processing, and 
(\emph{3}) analysis, done in this work~(\circled{8}, \Cref{fig:overview}).

For \textit{stage 1}, Lotito et al. at IBM collect \dataset~\cite{lotito26finetuning} with the SFTTrainer actuator of the open-source \texttt{ado} framework~\cite{DBLP:journals/jossw/JohnstonP26}, which runs fms-hf-tuning jobs~\cite{fms_hf_tuning} over a systematic sweep of configurations.
The dataset covers 33 LLMs of various scales~(0.35B-405B parameters), 3 fine-tuning methods, total batch sizes of \mbox{$2^0$-$2^{10}$}, and $2^9$-$2^{13}$ tokens per sample (sequence length).
The \textit{fine-tuning methods} are full~\cite{DBLP:conf/naacl/DevlinCLT19}, LoRA~\cite{DBLP:conf/iclr/HuSWALWWC22}, and GPTQ-LoRA~\cite{DBLP:conf/iclr/FrantarAHA23}, which applies LoRA to a GPTQ-quantized LLM.
The dataset also covers three \textit{optimizations}, whose impact we analyze:
(1)~Fast Kernels, fused low-level kernels~\cite{DBLP:conf/pldi/TilletKC19};
(2)~Fast \underline{MoE}, kernels of \underline{m}ixture-\underline{o}f-\underline{e}xperts with expert parallelism~\cite{DBLP:journals/corr/abs-2103-13262}; 
and (3)~\underline{RoCE}, \underline{R}DMA \underline{o}ver \underline{C}onverged \underline{E}thernet, which lets nodes exchange data directly between their memories, over Ethernet, without involving the CPU or the operating system~\cite{DBLP:conf/sigcomm/ZhuEFGLLPRYZ15}.
As \Cref{tab:background:dataset_stats} summarizes, for this dataset, LLMs run on 4 \textit{GPU types}: L40S as an enterprise GPU, A100 with two interfaces as a powerful GPU, and H100 as the expensive but fastest-in-class GPU.

\textit{(2) Data processing:} 
We keep all 30,920 \textit{experiments} from the dataset, unchanged. 
For each experiment, we extract its outcome (valid, rejected before launch, or failed at runtime), its batch size per GPU, %
and its optimization technique.

We infer from system policies external to the dataset that invalid experiments are rejected before launch if the batch size is not a multiple of the GPU count~(5,360 experiments); or, when the model uses parallel experts, if the GPU count is not a multiple of the \textit{expert-parallel~(EP)} degree~(3,784). %

\textit{(3) Data analysis:}
We define \textit{four key questions} a provider encounters when deploying LLM fine-tuning jobs (\Cref{tab:questions}), %
each with a metric and a target for analysis. %

\vspace*{-.3cm}
\section{LLM Fine-Tuning Characteristics}
\label{sec:analysis}

We follow the method introduced in \Cref{sec:method}, use the questions from \Cref{tab:questions}, and conduct a detailed characterization of LLM fine-tuning workloads running in GPU-based clusters. 
For each question, we begin with the most important findings, labeled \textbf{Finding i} for the \textbf{i}$^{\text{th}}$ finding, then present our detailed analysis. 

\subsection{(Q1) Failures: Will the Job Run?}
\label{sec:analysis:run}

Job failures, especially unrecoverable ones, delay execution, can cause queue bottlenecks, and, if failing at runtime, also waste GPU time, energy, and ultimately money~\cite{DBLP:conf/acl/StrubellGM19, DBLP:conf/ccgrid/KondoJIE10}.
We compute the failure ratio across four factors and illustrate the results in \Cref{fig:failures}. All failures are unrecoverable. %

\finding{fnd:1}{63\% experiments yield no valid run. Batch size and sequence length strongly correlate with runtime failure.} %

\finding{fnd:2} {Failures follow two batch-related trends: small batches are mostly rejected before launch because they cannot be evenly split across GPUs, and large batches mostly fail at runtime, likely because they exceed GPU memory.}

\textit{Overall}, 19,539 experiments (63\%) yield no valid run, with 9,144 rejected before launch and 10,395 failed at runtime.
The high failure ratio indicates that na\"{i}ve configurations should not be used. 
However, the distribution of failures across batch size and sequence length, \Cref{fig:failures}a and~b, respectively, suggests that profiling can identify values for these factors where the failure likelihood is significantly lower. 

\textit{Failures vs.\ batch size:}
The failure ratio is U-shaped in the batch size: from 86\% at batch size 1, which are mostly rejected before launch, it falls to 42\% at 8, then rises to 95\% at 1,024.
Thus, a profiling approach efficiently offers, through sampling, %
insights into failure and success patterns.

\textit{Failures vs.\ sequence length:}
Similarly, failures grow with sequence length, but follow a logarithmic trend~(\Cref{fig:failures}b, vertical axis is power-of-2) rather than a U-shape, because both extreme batches (low or high) and extreme sequence lengths (high) tend to overflow GPU memory. 
However, the main growth with higher sequence length comes from runtime failures, which increase from 21\% to 46\% of experiments. The experiments rejected before launch stay within 26-31\%, because the rules of the job runner (\textit{actuator}) in the IBM experiments do not involve sequence length~\cite{DBLP:journals/jossw/JohnstonP26}.

\textit{Failures vs.\ GPU type:}
When computational resources are scarce, feasibility under low memory becomes a primary concern alongside performance degradation~\cite {lotito26finetuning}. The dataset shows that the 48\,GB L40S has the highest failure ratio (69\%, \Cref{fig:failures}c) in 739 of its 1,068 experiments, mostly at runtime (54\%), while A100-PCIe, with 80\,GB, shows 58\% failures in 1,050 of its 1,818 experiments, consistent with its larger memory, though model mixes differ.

\textit{Failures vs.\ fine-tuning method:}
Methods that need less memory lower hardware requirements~\cite{DBLP:conf/iclr/HuSWALWWC22}; full fine-tuning, the most memory-hungry, fails 67\% of runs (\Cref{fig:failures}d), while LoRA and GPTQ-LoRA train only a fraction of the parameters and fail less often. 
At runtime, LoRA fails less often than full fine-tuning (30\% vs. 36\%), consistent with lower memory needs for LoRA.
However, memory is not always the cause; job distribution and the hardware used also matter: GPTQ-LoRA, for example, fails most at runtime (40\%), yet ran only on A100-SXM.

\textit{Implications:}
Automatic pre-launch checks can alleviate both classes of failures: rejection before launch and runtime failure.
A rule check for feasibility can repair a configuration, instead of rejecting it, for example by adjusting the batch size to a multiple of the number of GPUs.
A memory check can limit the batch size and sequence length based on available GPU memory, helping prevent runtime failure~\cite{lotito26finetuning}.

\subsection{(Q2) Performance: How Fast Will the Job Run?}
\label{sec:analysis:fast}

\begin{figure}[t]
  \centering
  \includegraphics[width=\columnwidth]{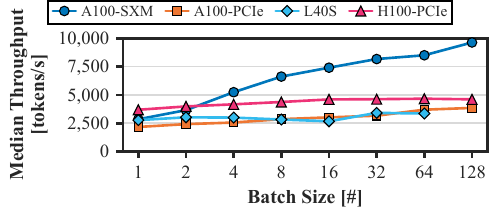}
  \vspace*{-0.75cm}
  \caption{Throughput of fine-tuning workloads deployed on a single-GPU cluster, by GPU type.}
  \Description{Throughput of fine-tuning workloads by GPU type.}
  \label{fig:batch}
\end{figure}

\begin{figure}[t]
\vspace*{-0.13cm}
  \centering
  \hspace*{-0.2cm}\includegraphics[width=1.08\columnwidth]{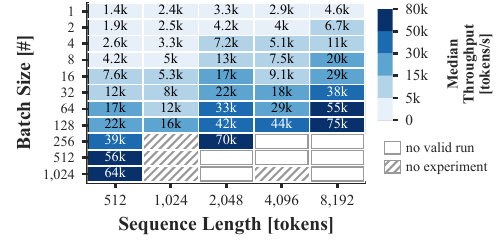}
  \vspace*{-0.9cm}
  \caption{Throughput of all 11,381 valid runs, per total batch size (rows) and sequence length (columns). Empty cells: no valid run; hatched: not tested.}
  \Description{Heatmap of the median throughput per batch size and sequence length, with empty cells where no run was valid.}
  \label{fig:heatmap}
  \vspace*{-0.4cm}
\end{figure}

Throughput directly influences how many GPU hours a fine-tuning job costs and the derived metrics of the job (e.g., energy consumption, financial cost, cluster usage, time in queue).
The dataset measures throughput as the number of tokens a job processes per second; for the batch-size trend, to exclude the effect of adding GPUs, we use only the 2,320 valid runs on a single GPU.

\finding{fnd:3}{Throughput grows logarithmically with the batch size: on one A100-SXM GPU, batch size 8 vs.\ 1 increases throughput by 2.3$\times$, yet batch 128 (vs.\ 8) increases throughput only by 1.5$\times$.}

\textit{Throughput vs.\ batch size:}
\Cref{fig:batch} illustrates a logarithmic relationship between throughput and batch size: each doubling of the batch size adds approximately the same amount, about 1,000 tokens/s on A100-SXM.
For this GPU, the throughput rises from 2,847 tokens/s at batch 1 to 6,616 at 8, but only to 9,627 at 128, where the memory starts to saturate; beyond batch size 128, no single-GPU run succeeded.

\textit{Throughput vs.\ GPU type:}
Because each GPU type ran different models, we can compare GPU types only on paired experiments where the only variable is the GPU type.
In all 103 pairs, the 48\,GB L40S is slower than the 80\,GB A100-PCIe, with a median throughput 32\% lower.
The A100-SXM, in turn, is 18\% faster than A100-PCIe in all 49 pairs.
However, we cannot pair H100-PCIe because the dataset has no identical job on H100-PCIe and another GPU type.

\textit{Throughput vs.\ sequence length:}
Throughput mostly grows with both batch size and sequence length, as \Cref{fig:heatmap} shows, from 1.4k tokens/s at (batch size, sequence length) of (1, 512) to 75k at (128, 8\,192) on 16 or 32 GPUs.
However, increasing further leads to infeasible, failing jobs.

\textit{Implications:}
Failed or only narrowly faster runs are expensive, which affects small budgets (e.g., academic groups)~\cite{DBLP:conf/acl/StrubellGM19}.
Large batches only narrowly increase throughput but raise failure risk (\reffinding{fnd:1}), leaving moderate batches as the all-around choice.%

\begin{figure}[t]
  \centering
  \includegraphics[width=\columnwidth]{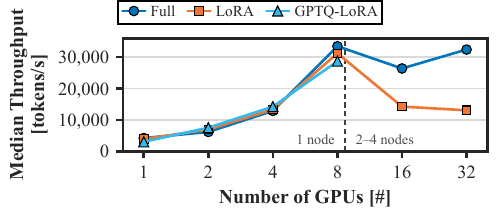}
  \vspace*{-0.75cm}
  \caption{GPU scaling per workload (1 node = 8 GPUs).}
  \Description{GPU scaling per workload (1 node = 8 GPUs).}
  \label{fig:scaling}
  \vspace*{-0.4cm}
\end{figure}

\subsection{(Q3) Scalability: Will the Job Scale with GPUs?}\label{sec:analysis:gpus}\label{sec:analysis:scalability}

The cost of more GPUs is worth it only if throughput amortizes it, and the extra GPUs do not sit idle or waste time communicating~\cite{DBLP:journals/access/LotitoVVPPJ26}.
\Cref{fig:scaling} shows median throughput versus the number of GPUs (1 to 32) across homogeneous nodes with 8 GPUs each; up to 8 GPUs, the jobs run intra-node and, otherwise, inter-node.
Because different LLMs ran at different GPU counts, we also analyze the weak-scaling efficiency on matched pairs, jobs that differ only in the number of GPUs and keep the same batch size per GPU (efficiency of 1 means that doubling the GPUs also doubles the throughput).

\finding{fnd:4}{Intra-node, throughput scales almost linearly with the number of GPUs (efficiency of 0.94-1.00). Across nodes, scaling is also influenced by the run setup and the network: from 8 to 16 GPUs (1 to 2 nodes), efficiency is 0.98 for runs without a time limit, but drops to 0.26 without RoCE and to 0.72 with RoCE for runs capped at 600\,s.}

\textit{Within a node:}
Up to 8 GPUs, all three methods scale near-linearly: median throughput grows from 3,100-4,300 tokens/s on one GPU to 28,700-33,500 on 8, and matched efficiency stays between 0.94 and 1.00 for every intra-node doubling.
With identical total batch (strong scaling), efficiency drops to only 0.85-0.88 (104-207 pairs), mainly because individual GPUs receive less work.

\textit{Across nodes:}
From 8 to 16 GPUs, thus, on 2 nodes, GPUs also need to communicate over the network, which can affect efficiency.
However, runs without a time limit (fms-hf-tuning 2.4.0) gain 1.95-1.96$\times$ throughput with and without RoCE,
while runs capped at 600\,s (fms-hf-tuning 2.7.1) gain only 1.45$\times$ with RoCE.
The same 600\,s-capped jobs, without RoCE, slow to 0.52$\times$, equivalent to doubling the compute power and halving the throughput.
From 16 to 32 GPUs (2 to 4 nodes), the same 600\,s-capped jobs gain 1.88$\times$ without RoCE and only 1.90$\times$ with it, showing that when jobs already run distributed across nodes, adding more nodes brings close to linear gains, with efficiency at 0.94-0.95.

\textit{Scaling vs.\ fine-tuning method:}
Up to 8 GPUs, the three methods scale alike;
above 8 GPUs, the LoRA curve from \Cref{fig:scaling} drops below full fine-tuning because the LoRA runs used much larger models (median 47B parameters) than the full fine-tuning runs (8.1B).
However, when the same job runs with both methods, LoRA is only 9\% faster than full fine-tuning (median for 3,744 pairs).

\textit{Implications:}
Jobs up to 8 GPUs should run within one node, where doubling the GPUs also (nearly) doubles the throughput.
The inter-node world is less predictable, with gains depending on the run setup and network: without RoCE, some jobs ran at half their one-node speed even when using twice the GPUs.

\begin{table*}[t]
  \newcommand{\hbfull}{\faCircle}%
  \newcommand{\hbhalf}{\faAdjust}%
  \newcommand{\hbnone}{\faCircle[regular]}%
  \caption{Actionable insights: techniques and candidate systems that could use the findings of \Cref{sec:analysis}: (\textbf{F1})~U-shaped, batch-driven failure ratio, (\textbf{F2})~batch-dependent pre-launch and runtime failures, (\textbf{F3})~logarithmic batch-size throughput growth, (\textbf{F4})~near-linear intra-node, variable inter-node scaling, (\textbf{F5})~setup-dependent system-level RoCE speedups, and (\textbf{F6})~application-level MoE and kernel speedups. Each finding informs a technique directly (\hbfull), partly (\hbhalf), or not (\hbnone).}
  \label{tab:insights}
  \vspace*{-0.3cm}
  \footnotesize
  \centering
  \setlength{\tabcolsep}{2pt}
  \setlength{\aboverulesep}{0pt}\setlength{\belowrulesep}{0pt}%
  \renewcommand{\arraystretch}{1.12}
  \begin{tabular}{|l|l|l|*{6}{@{\hspace{0.7pt}}c@{\hspace{0.7pt}}|}l|l|}%
    \toprule
     & & & \multicolumn{6}{c|}{\textbf{Finding}} & & \\
    \cline{4-9}
    \textbf{Domain} & \textbf{Technique} & \textbf{Key evidence} & \textbf{1} & \textbf{2} & \textbf{3} & \textbf{4} & \textbf{5} & \textbf{6} & \textbf{Potentially benefiting systems} & \textbf{Objective} \\
    \midrule
    Operations & Admission control & 63\% of experiments fail & \hbfull & \hbfull & \hbhalf & \hbnone & \hbnone & \hbnone & AutoConf~\cite{ibm_autoconf}, Ymir~\cite{DBLP:conf/ics/GaoZLS0024} & Fewer failures \\
    & Right-sizing & Batch 8$\rightarrow$128: only 1.5$\times$ throughput & \hbhalf & \hbhalf & \hbfull & \hbhalf & \hbnone & \hbnone & Morph.~\cite{DBLP:conf/cloud/WangYYWLSHZ21}, \coastline~(\S\ref{sec:conclusion})~\cite{nicolae26coastline} & Resource use \\
     & Placement & Intra-node efficiency 0.94-1.00 & \hbnone & \hbnone & \hbnone & \hbfull & \hbhalf & \hbnone & FGD~\cite{DBLP:conf/usenix/WengYY0TYZ23}, AlpaServe~\cite{DBLP:conf/osdi/LiZZL00HCZGS23} & Resource use, latency \\
     & Optimization choice & Speedups up to 2.9$\times$ & \hbnone & \hbnone & \hbnone & \hbhalf & \hbfull & \hbfull & INFaaS~\cite{DBLP:conf/usenix/Romero0YK21}, \coastline~\cite{nicolae26coastline} & Throughput \\
     & Provisioning & 11-65\% fewer GPU hours & \hbhalf & \hbnone & \hbhalf & \hbhalf & \hbfull & \hbfull & SpotServe~\cite{DBLP:conf/asplos/MiaoSDXL0J24} & Cost-efficiency \\
    \midrule
    Workflow analysis & Tracing & Throughput and failure profiles & \hbhalf & \hbhalf & \hbfull & \hbnone & \hbnone & \hbnone & TraceIT~\cite{DBLP:journals/fgcs/ScheunerETEALI26} & Timing reconstruction \\
    \midrule
    Simulation & What-if models & Matched pairs as ground truth & \hbhalf & \hbhalf & \hbfull & \hbfull & \hbfull & \hbfull & OpenDC~\cite{DBLP:conf/ccgrid/MastenbroekAJLB21}, \kavier~\cite{DBLP:journals/corr/abs-2605-25247} & Long-term planning \\
    \midrule
    Digital twinning & Online recalibration & Calibration data & \hbfull & \hbfull & \hbfull & \hbfull & \hbfull & \hbfull & OpenDT~\cite{DBLP:conf/wosp/NicolaeTKLI26}, ExaDigiT~\cite{DBLP:conf/sc/BrewerMKWBHSGGW24} & Live operations \\
    \bottomrule
  \end{tabular}
  \vspace*{-0.2cm}
\end{table*}

\subsection{(Q4) Optimization: Faster Fine-Tuning?}
\label{sec:analysis:faster}

\begin{figure}[t]
  \centering
  \includegraphics[width=\columnwidth]{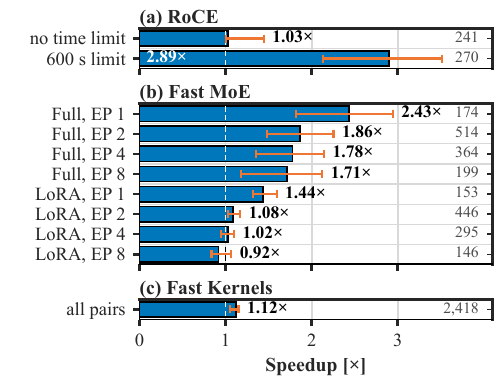}
  \vspace*{-0.7cm}
  \caption{Median speedup over matched pairs of jobs that differ only in one optimization: (a)~RDMA over Converged Ethernet, (b)~faster code for MoE layers, or (c)~faster code for common training operations. EP: expert-parallel degree; whiskers: middle 50\%; gray: number of pairs.}
  \Description{Optimization ``return on investment''.}
  \vspace*{-0.7cm}
  \label{fig:speedups}
\end{figure}

Fine-tuning optimizations, such as LoRA, can lower the already high multi-tier cost associated with LLM workloads~\cite{DBLP:journals/access/LotitoVVPPJ26, DBLP:conf/iiswc/XiaKCY0HT24}.
\dataset profiles three optimizations (\Cref{sec:method}): 
(a) RoCE, a system-level optimization that speeds up the data exchange between nodes~\cite{DBLP:conf/sigcomm/GuoWDSYPL16};
(b) Fast MoE, an application-level optimization that speeds up the expert layers of MoE models~\cite{fms_acceleration};
and (c) Fast Kernels, an application-level optimization that replaces the standard GPU code of frequent training computations (e.g., loss, normalization layers) with faster, hand-optimized code~\cite{DBLP:conf/pldi/TilletKC19, fms_acceleration}.
\Cref{fig:speedups} depicts speedups of each optimization.

\finding{fnd:5}{\textbf{At system level,} RoCE speeds up multi-node jobs, especially runs capped at 600\,s: 2.89$\times$, vs.\ 1.03$\times$ without a time limit.}

\finding{fnd:6}{\textbf{At application level,} Fast MoE speeds up MoE LLMs by 1.69$\times$ when all experts stay single-GPU, but only 1.16-1.33$\times$ when the experts spread over 2, 4, and 8 GPUs. Fast Kernels, however, add only 12\% throughput (1.12$\times$).}

\textit{System-level, RoCE (\Cref{fig:speedups}a):}
Across 511 job pairs on 2-4 nodes, RoCE speeds up fine-tuning by a median of 2.16$\times$.
This median, however, mixes two classes of runs with contrasting results.
Runs without a time limit gain only 3\% throughput~(1.03$\times$, 241 pairs); on 2 nodes, they gain a mere 1\%.
Runs with a 600\,s time limit gain 2.89$\times$ (a quarter of them above 3.5$\times$).
The two classes also differ in gradient accumulation (mostly 4 vs.\ 1).

\textit{Fast MoE (\Cref{fig:speedups}b):}
MoE models send each token only through a small fraction of their expert sub-networks~\cite{DBLP:journals/corr/abs-2401-04088}, and Fast MoE runs these experts with hand-optimized GPU code~\cite{DBLP:journals/corr/abs-2403-08245, fms_acceleration}.
When the experts stay on 1 GPU (EP=1), Fast MoE speeds up fine-tuning by a median of 1.69$\times$: 2.43$\times$ for full fine-tuning and 1.44$\times$ for LoRA. The speedup gap arises because LoRA leaves the expert weights frozen, as follows from~\cite{DBLP:conf/iclr/HuSWALWWC22}, and there is less expert computation left to speed up.
When the experts spread over GPUs (EP>1), we identify a decreasing trend, also illustrated in \Cref{fig:speedups}b; for example, for LoRA at EP=8, speedup even becomes a slowdown (0.92$\times$), because exchanging tokens between GPUs, which expert parallelism requires~\cite{DBLP:journals/corr/abs-2103-13262}, takes time the faster code cannot amortize.
The higher the EP, the more communication dominates, and slowdowns occur when latency hiding is no longer feasible.

\textit{Fast Kernels (\Cref{fig:speedups}c):}
They have a smaller but steady speedup of 1.12$\times$ over 2,418 pairs of seven models of up to 8B parameters. Half of these pairs gain 5-15\%.

\textit{Implications:}
We identify that the optimizations mostly save GPU hours (hence, energy and money): 11\% saved for Fast Kernels, 65\% for RoCE in runs capped at 600\,s, but also minor losses such as LoRA at EP=8 (0.92$\times$, a slowdown).
Based on these findings, a provider can enable Fast Kernels by default, use Fast MoE for LLM experts that fit on one GPU, and enable RoCE for multi-node jobs after checking that jobs benefit (in the dataset, jobs with a 600\,s limit). 

\section{Actionable Insights}
\label{sec:insights}

In this section, we present actionable insights that help achieve the vision formulated in the introduction: A serverless infrastructure for LLM fine-tuning, where profiling data and resulting characterization help achieve better configurations.
\Cref{tab:insights} links our main findings from \Cref{sec:analysis}
with four domains to improve toward our vision. For each domain, we identify techniques and specific objectives to be improved by leveraging the key evidence produced in this work. 
We also identify concrete examples of systems~(tools) that can help with the objectives. We now address each domain. %

\textit{Operations:}
\dataset and the correlated findings can be used as input to improve dynamic decision-making, resource management, and scheduling to improve 
latency~\cite{DBLP:conf/osdi/LiZZL00HCZGS23} or throughput~\cite{DBLP:conf/usenix/Romero0YK21},
resource use~\cite{DBLP:conf/cloud/WangYYWLSHZ21,DBLP:conf/usenix/WengYY0TYZ23}, and
cost-efficiency~\cite{DBLP:conf/asplos/MiaoSDXL0J24}.
For fine-tuning, analysis of the feasibility boundary (e.g., our F1, F2) or batch sizes (e.g., F3) can guide analysis-driven admission control in tools such as AutoConf~\cite{ibm_autoconf} or Ymir~\cite{DBLP:conf/ics/GaoZLS0024}. 
Accessibility is also rapidly becoming an operational concern: The ongoing GPU scarcity limits accessibility~\cite{DBLP:journals/access/LotitoVVPPJ26}; moreover, accessible units across the continuum could be subject to specific licensing, regulations, and other concerns that restrict operation~\cite{DBLP:conf/ccgrid/JansenAPTI23}. 

\textit{Workflow analysis:}
Our findings on throughput and failure profiles per configuration (F1-F3) can also be used 
to enable performance analysis for complex serverless workflows, when combined with other tracing and profiling techniques, such as TraceIT~\cite{DBLP:journals/fgcs/ScheunerETEALI26} for timing reconstruction when asynchronous and third-party services are part of the workflow.

\textit{Simulation:}
The matched pairs (F4-F6) can serve as measured and processed reality and can be 
used as input for complex simulations~\cite{DBLP:conf/ccgrid/MastenbroekAJLB21,DBLP:journals/tpds/TalluriNCKCBI26} for long-term analysis and optimization processes, for example with \kavier~\cite{DBLP:journals/corr/abs-2605-25247}, to evaluate resource management and scheduling policies for performance, capacity planning, cost-efficiency, but also for the more emerging sustainability concerns; we envision such analysis, prior to deployment, to become a state-of-the-art procedure.
Simulators such as \kavier~\cite{DBLP:journals/corr/abs-2605-25247} could leverage \dataset to derive analytical or ML-based models to predict LLM fine-tuning performance, or to calibrate existing models using the measured reality.

\textit{Digital twinning:}
As an overarching goal, the IBM dataset can be used as input for digital twinning of live operations~\cite{DBLP:conf/wosp/NicolaeTKLI26}, leveraging all the aforementioned classes of actions benefiting from \dataset: (1) operations, (2) workflow analysis, or (3) simulation.
A digital twin could leverage \dataset as calibration data or as a ground truth (better: measured reality) for the twin to use when making decisions on infrastructure steering, either automated or with a human-in-the-loop~\cite{DBLP:conf/wosp/NicolaeTKLI26}. %
Such a digital twin, applied to LLM fine-tuning workloads, could use \kavier~\cite{DBLP:journals/corr/abs-2605-25247} as its simulation engine and \coastline~\cite{nicolae26coastline} as its advisor, with both helping steer serverless infrastructure toward established service level objectives (SLOs) and agreements (SLAs).

\section{Conclusion and Ongoing Work}
\label{sec:conclusion}

Serverless LLM fine-tuning is an increasingly common workload class, which transfers the complexity of job configurations to providers. However, only scarce empirical evidence exists on how these workloads deploy, run, and succeed. %
Using the public dataset \dataset released by IBM, in this work we characterize LLM fine-tuning workloads in GPU-based clusters.
We find that 
(1)~failures, frequent due to the sweep design, follow two batch-dependent modes, 
(2)~throughput only grows logarithmically with the batch size, 
(3)~scaling is near-linear within a node, but setup-dependent across nodes, and 
(4)~optimizations such as Fast Kernels, RoCE, or Fast MoE could yield 
significant
speedups.
Based on our findings, we derive actionable insights (\S\ref{sec:insights}).

\textbf{Ongoing work:} 
Toward reliable, fast, and efficient LLM fine-tuning, we pursue three directions of increasing magnitude.
First, in the \textit{operations} domain, we are integrating the feasibility boundary (Findings~1-2) and observed performance trends (Findings~3-6) into~\coastline~\cite{nicolae26coastline}, a system
that helps users configure fine-tuning jobs.

For \textit{simulators}, we are contributing to~\kavier~\cite{DBLP:journals/corr/abs-2605-25247}, an analytical discrete-event simulator, by expanding its capabilities with fine-tuning, calibrated and validated against\\ \dataset, which serves as measured ground truth (Findings~3-6). %

Last, for \textit{digital twins}, we envision that a digital counterpart of the real-world ecosystem could steer toward objective-oriented outcomes to better meet established SLOs and SLAs. 
Integrating the operations and simulator tools described earlier could help achieve this vision. Furthermore, this is a call for the entire community to help develop meaningful and versatile digital twins that can become trusted advisors-on-the-side in serverless LLM operations.

\begin{anonsuppress}
\begin{acks}
This work was partially supported by EU MSCA AIStars4EU (g.a. 101341806) and partly supported by a National Growth Fund through the Dutch project 6G Future Network Services.
\end{acks}
\end{anonsuppress}

\bibliographystyle{ACM-Reference-Format}
\bibliography{references}

\end{document}